\documentclass{article}
\usepackage{graphicx}
\graphicspath{%
    {converted_graphics/}
    {/}
}
\begin{document}

\begin{center}
{\LARGE Measurements of Growth Rates of (0001) Ice Crystal Surfaces}\vskip6pt

{\Large K. G. Libbrecht and M. E. Rickerby}\vskip4pt

{\large Department of Physics, California Institute of Technology}\vskip-1pt

{\large Pasadena, California 91125}\vskip-1pt

\vskip18pt

\hrule\vskip1pt \hrule\vskip14pt
\end{center}

\textbf{Abstract}. We present measurements of growth rates of the (0001)
facet surface of ice as a function of water vapor supersaturation over the
temperature range $-2$ $\geq T\geq -40$ C. From these data we infer the
temperature dependence of premelting on the basal surface and the effects of
premelting on the ice growth dynamics. Over this entire temperature range
the growth was consistent with a simple 2D nucleation model, allowing a
measurement of the critical supersaturation $\sigma _{0}(T)$ as a function
of temperature. We find that the 2D nucleation barrier is substantially
diminished when the premelted layer is partially developed, as indicated by
a reduced $\sigma _{0},$ while the barrier is higher both when the premelted
layer is fully absent or fully developed.

\section{Introduction}

It has long been long suspected that premelting in ice plays an important
role in the growth dynamics of ice crystals from water vapor \cite%
{kurodalac, kkreview, libbrechtreview}. Although ice is a monomolecular
crystal with a simple hexagonal structure under normal atmospheric
conditions, ice crystals forming from water vapor exhibit an exceedingly
rich spectrum of plate-like and columnar morphologies as a function of
temperature and supersaturation over the temperature range $0$ $\geq T\geq
-30$ C \cite{libbrechtreview}. Since the premelted layer in ice develops
over this same temperature range \cite{dash, wei, dosch}, the prevailing
thinking holds that temperature-dependent effects of premelting on ice
crystal growth are responsible for the observed morphological complexities,
together with instabilities arising from diffusion-limited growth and other
effects \cite{libbrechtreview}. To date, however, this long-held hypothesis
has remained largely unsupported by solid experimental evidence.

We sought to shed light on this problem by making precise measurements of
the growth rates of small faceted ice crystals from water vapor under
carefully controlled conditions, in order to better quantify the intrinsic
ice growth behavior. To this end we measured growth rates of (0001) ice
surfaces as a function of water vapor supersaturation $\sigma $ and
temperature $T$ over the temperature range $-2$ $\geq T\geq -40$ C. Our
measurements were made at low background pressure to reduce the effects of
particle diffusion, so the growth was mainly limited by attachment kinetics.

As we will show below, our measurements were all consistent with a simple 2D
nucleation model of crystal growth, and from our data we extracted the
critical supersaturation $\sigma _{0}(T)$ as a function of temperature. The
function $\sigma _{0}(T)$ showed an interesting behavior from which we were
able to infer the onset and development of premelting on the ice surface as
well as how premelting affects crystal growth in ice.

\section{Ice Crystal Growth Measurements}

\begin{figure}[ht] 
  \centering
  \includegraphics[width=4.3in,keepaspectratio]{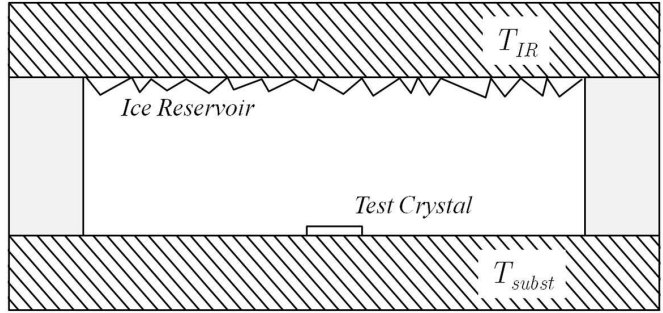}
  \caption{An idealized schematic of our
experimental set-up. The top surface is an ice reservoir at temperature $%
T_{IR}$ that supplied water vapor for a test crystal resting on a substrate
at temperature $T_{subst}.$ When $T_{IR}>T_{subst}$, growth rates were
determined by measuring the size and thickness of the test crystal as a
function of time.}
  \label{basic}
\end{figure}

The goal
of our ice growth experiments was to examine the growth of individual ice
crystals in a carefully controlled environment, and an idealized schematic
diagram of our experimental set-up is shown in Figure \ref{basic}. The top
surface of the experimental chamber was a thermal conductor with a uniform
temperature $T_{IR},$ and its inside surface was covered with a layer of ice
crystals that made up an ice reservoir. At the beginning of each
measurement, a single test crystal was placed near the center of the bottom
substrate surface held at temperature $T_{subst}$. The walls were thermally
insulating, and the vertical spacing from the top to the bottom of the
chamber was 1.0 mm. The temperature difference $\Delta T=T_{IR}-T_{subst}$
determined the effective supersaturation seen by the test crystal. During
the course of a measurement we increased $\Delta T$ and observed the size
and thickness of the test crystal as a function of time, and from this we
extracted crystal growth velocities as a function of supersaturation. A
detailed hardware description is provided in \cite{details}. Some important
aspects of the experiment included:

1) Our test crystals were small, thin plates, typically $<5$ $\mu $m thick
and $<50$ $\mu $m in diameter. This was important to reduce the effects of
particle and heat diffusion, as described quantitatively in \cite{details}.

2) We used crystals with one basal facet in contact with the substrate,
measuring the perpendicular growth of the opposite (0001) surface using
broad-band interferometry as described in \cite{details}. Growth of the
prism facets was affected by substrate interactions, particularly at low
supersaturations, so these measurements were discarded.

3) We used only crystals with simple morphologies and well formed facets,
and each crystal was discarded after growth. Evaporating and regrowing
crystals was found to result in generally lower quality data.

4) Our test crystals were freshly made in a clean environment and
transported within minutes to our test chamber with minimal processing, as
described in \cite{details}.

With these precautions, and exercising considerable care to create a stable
growth environment with precisely known supersaturations, we were able to
obtain quite satisfactory data, and an example is shown in Figure \ref%
{basalgrowth}. Here we have plotted the effective attachment coefficient $%
\alpha $, which is derived from the perpendicular growth velocity $v$ using $%
v=\alpha v_{kin}\sigma ,$ where $\sigma $ is the supersaturation far from
the crystal and $v_{kin}$ is a kinetic velocity \cite{libbrechtreview,
details}.

\begin{figure}[ht] 
  \centering
  \includegraphics[width=4.25in,height=3.26in,keepaspectratio]{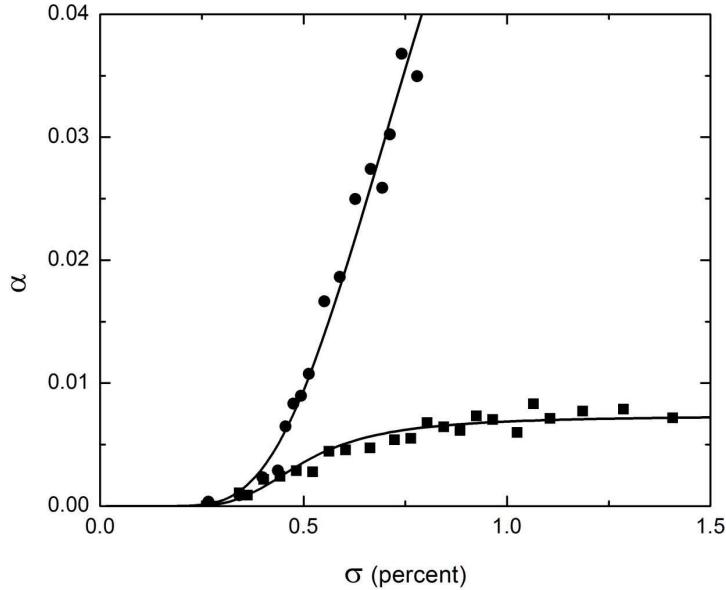}
  \caption{Measurements of the growth of
the basal facets of two ice crystals at -15 C, shown as the effective
condensation coefficient $\protect\alpha $ as a function of supersaturation $%
\protect\sigma $ far from the crystal. One crystal (dots) was grown in a
background pressure of air at 20 Torr and the other (squares) was grown in a
background pressure of 740 Torr. The low-pressure crystal shows mainly
kinetics-limited growth, while the growth at high pressure is mainly limited
by particle diffusion when the supersaturation is high. Fit lines are
described in the text.}
  \label{basalgrowth}
\end{figure}

For our main data set, each crystal was grown at a pressure near 20 Torr,
where the growth was predominantly limited by attachment kinetics, at least
for small crystals at low supersaturations. For essentially all our data, we
found that the basal growth was well described by a 2D nucleation model \cite%
{saito}, and to describe the growth we adopted a simplified parameterization
of the intrinsic attachment coefficient $\alpha _{intrinsic}(\sigma
,T)=A\exp (-\sigma _{0}/\sigma )$ where $A$ and $\sigma _{0}$ are parameters
that may depend on temperature but not on supersaturation.

At high supersaturations our data were distorted by the effects of particle
and heat diffusion, so we modeled our data by fitting to the functional form%
\begin{equation}
\alpha \left( \sigma \right) =\frac{A\exp \left( -\sigma _{0}/\sigma \right)
\alpha _{fit}}{A\exp \left( -\sigma _{0}/\sigma \right) +\alpha _{fit}}
\label{fitform}
\end{equation}%
as described in \cite{details}. To produce more stable fits in our analysis,
we adjusted $\sigma _{0}$ and $\alpha _{fit}$ while keeping the scale factor 
$A$ fixed, and we chose $A=1$ to give the physically reasonable result that $%
\alpha _{intrinsic}\rightarrow 1$ as $\sigma $ becomes large. In Figure \ref%
{basalgrowth}, for example, the fit parameters are $(A,\sigma _{0},\alpha
_{fit})$ = $(1,2.3,0.15)$ and $(1,2.5,0.0075)$ for the low-pressure and
high-pressure crystals, respectively. As described in \cite{details}, the
value of $\alpha _{fit}$ depended mainly on the background pressure as well
as the size and thickness of each test crystal, so this parameter was of
little physical interest in our later analysis; the main focus in this
experiment was on measuring the critical supersaturation $\sigma _{0}$ as a
function of temperature.

\begin{figure}[htbp] 
  \centering
  \includegraphics[bb=0 0 999 1471,width=4.6in,keepaspectratio]{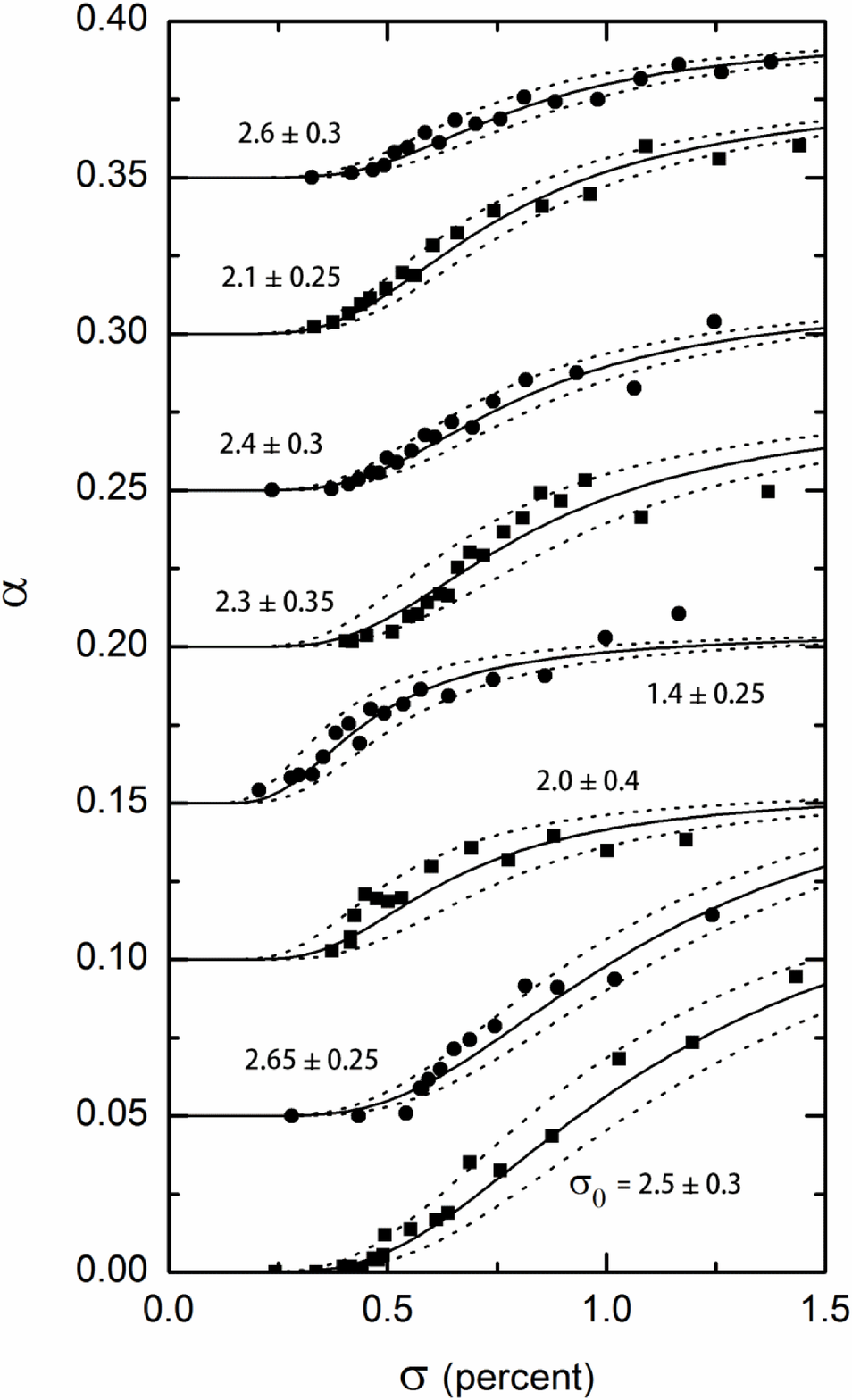}
  \caption{Eight separate data sets
showing the growth of crystals at -12 C. Individual measurements of $\protect%
\alpha \left( \protect\sigma \right) $ have been offset by multiples of 0.05
for clarity; the amount of each offset can be seen by noting that $\protect%
\alpha \left( 0\right) =0$. Solid lines drawn through each data set show
fits to the functional form described by Equation \protect\ref{fitform}, and
each curve is labeled with the fit critical supersaturation $\protect\sigma %
_{0}.$}
\label{data12a}
\end{figure}

Figure \ref{data12a} shows data from a typical six-hour run. Although each crystal
took only about ten minutes to measure, finding suitable crystals added
time, plus there was additional procedural overhead, so the experiment ended
up taking about 45 minutes per crystal. For each measurement, like those
shown in Figure \ref{data12a}, we generated curves bracketing the fit curves
as a visual means of producing an error estimate $\delta \sigma _{0}$ for
each measured $\sigma _{0}$. The dotted lines in Figure \ref{data12a} show
the fit curves for each data set except with $\sigma _{0}$ replaced by $%
\sigma _{0}\pm \delta \sigma _{0}$ in the fitting function Equation \ref%
{fitform}. In total, 102 crystals were measured and fit in this way to
produce our primary reduced data set.

We combined the individual $\sigma _{0,i}$ measurements at each temperature
by weighting each measurement with our estimated $\delta \sigma _{0,i}$ for
that measurement, thus producing a weighted mean $\left\langle \sigma
_{0}\right\rangle $ at each temperature along with an uncertainty estimate $%
\delta \left\langle \sigma _{0}\right\rangle $ for the mean \cite{taylor}%
\begin{eqnarray*}
\left\langle \sigma _{0}\right\rangle &=&\frac{\sum \sigma _{0,i}\delta
\sigma _{0,i}^{-2}}{\sum \delta \sigma _{0,i}^{-2}} \\
\delta \left\langle \sigma _{0}\right\rangle ^{2} &=&\frac{\sum \left(
\sigma _{0,i}-\left\langle \sigma _{0}\right\rangle \right) ^{2}\delta
\sigma _{0,i}^{-2}}{(N_{eff}-1)\sum \delta \sigma _{0,i}^{-2}}
\end{eqnarray*}%
where $N_{eff}=$ $\left( \sum \delta \sigma _{0,i}^{-2}\right) /\delta
\sigma _{0,\min }^{-2}$ is the effective number of points in the sample at
each temperature.

We examined our data fitting and analysis practices carefully and reached
several conclusions: 1) Nearly all our data were well fit by Equation \ref%
{fitform}, consistent with a simple 2D-nucleation model distorted by
particle diffusion; 2) Roughly ten percent of the crystals sampled grew very
rapidly, with essentially no nucleation barrier \cite{precisiongrowth}.
These crystals may have had dislocations, and they were discarded from our
data. These crystals were obviously different from the norm, however, and we
do not believe this practice was detrimental to our analysis; 3) Our fits
were quite robust to the critical supersaturation $\sigma _{0},$ since this
term was determined mainly by the growth behavior at low $\sigma ,$ while
diffusion predominantly affected the high-$\sigma $ growth. Since our data
were taken at low pressure, the inclusion of $\alpha _{fit}$ mainly helped
fit the higher-$\sigma $ part of the each curve, which was useful for
producing a better global fit.

\begin{figure}[ht] 
  \centering
  \includegraphics[bb=0 0 1211 984,width=4.71in,height=3.83in,keepaspectratio]{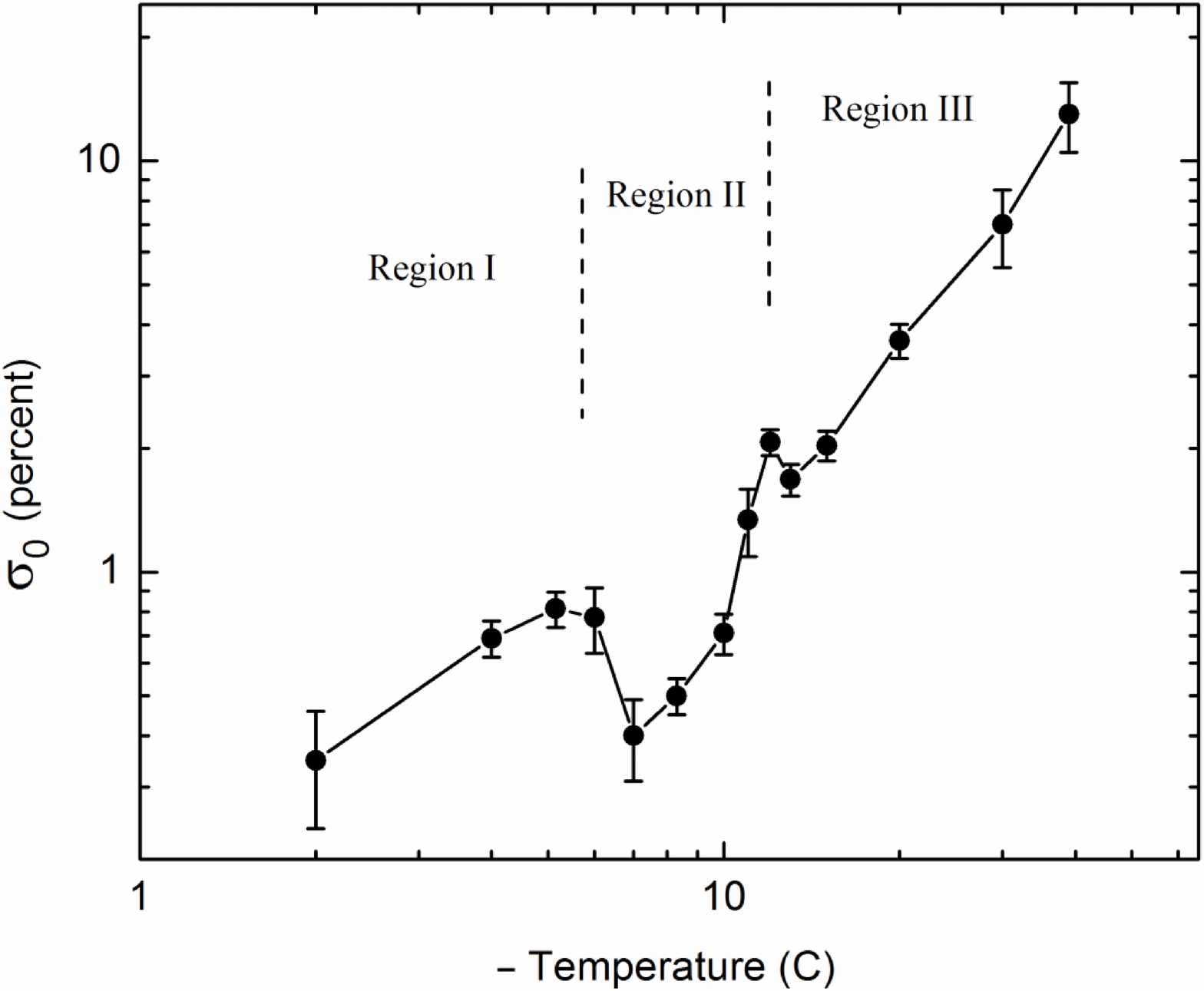}
  \caption{Combined measurements of the
critical supersaturation $\protect\sigma _{0}$ as a function of temperature,
as described in the text.}
  \label{finalgraph}
\end{figure}

Averaging our data at each temperature yielded the final measurement of the
critical supersaturation as a function of temperature $\sigma _{0}(T)$ shown
in Figure \ref{finalgraph}, which is the principal result from our
experiment. Here the current set of measurements includes only data points
for $T\geq -20$ C; the last two data points in Figure \ref{finalgraph} were
taken from \cite{oldgrowth}, which was a previous version of the present
experiment.

\begin{figure}[ht] 
  \centering
  \includegraphics[bb=0 0 1211 929,width=4.9in,height=3.76in,keepaspectratio]{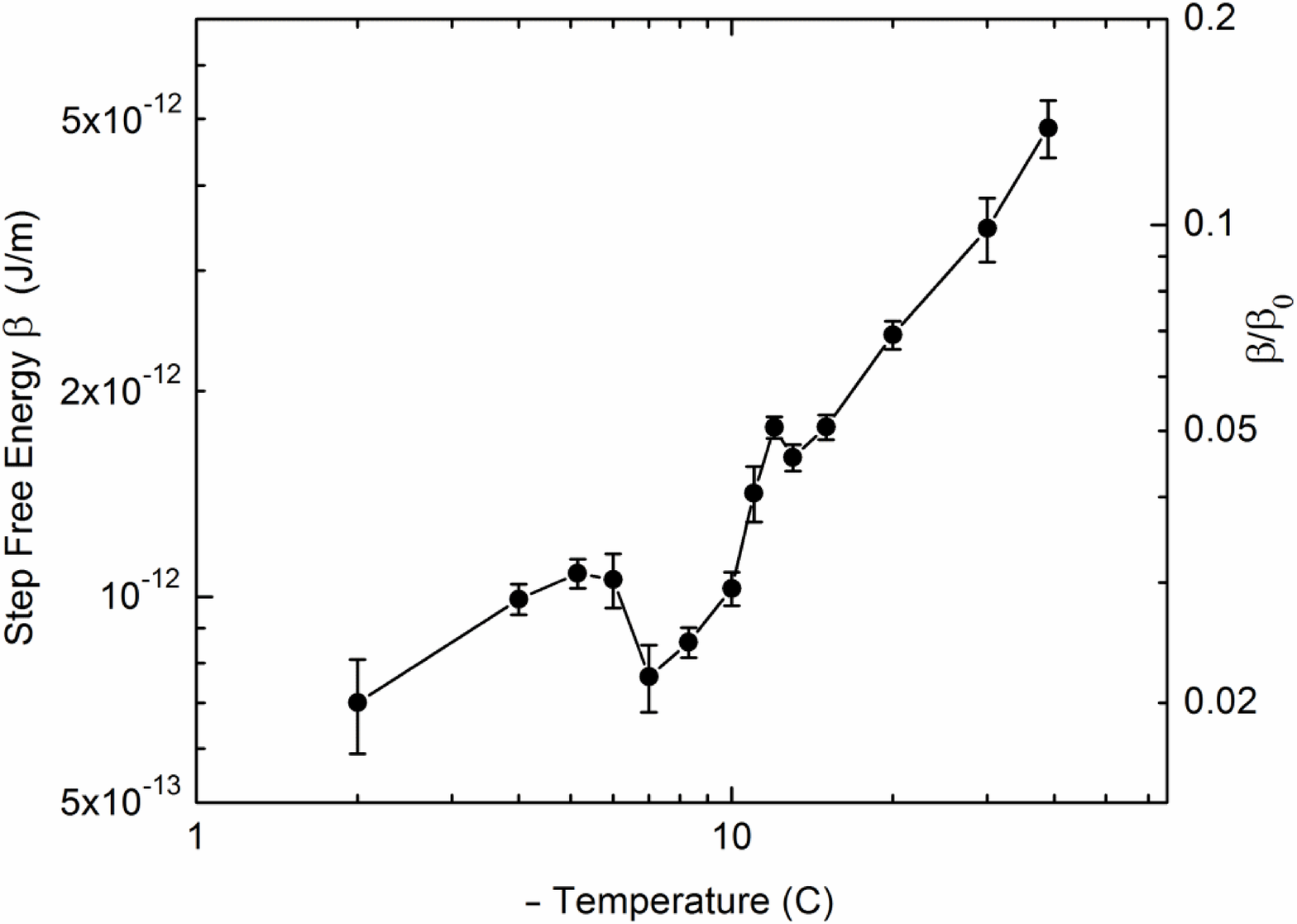}
  \caption{Step free energy $\protect\beta %
(T)$ as a function of temperature, calculated from the critical
supersaturation $\protect\sigma _{0}(T)$ according to classical nucleation
theory, as described in the text. The scale on the right side of the graph
shows $\protect\beta (T)$ normalized by $\protect\beta _{0},$ the product of
the surface energy and the step height \protect\cite{oldgrowth}.}
  \label{betagraph}
\end{figure}

Since the crystal growth we observed was everywhere well described by a 2D
nucleation model, the measured critical supersaturation $\sigma _{0}$ can be
used to calculate the step free energy $\beta $ using 
\[
\sigma _{0}=\frac{\pi \beta ^{2}\Omega _{2}}{3k^{2}T^{2}} 
\]%
where $\Omega _{2}$ is the area of a molecule on the surface, a relation
that comes from classical 2D nucleation theory \cite{saito, oldgrowth}. A
plot of $\beta (T)$ from our data is shown in Figure \ref{betagraph}. We
note from the scale on the right side of Figure \ref{betagraph} that $\beta
(T)$ is much smaller than $\beta _{0}=\gamma a\approx 3.5\times 10^{-11}$
J/m, the product of the surface energy $\gamma =0.11$ J/m$^{2}$ of the
ice/vapor interface and the nominal molecular step height $a=0.32$ nm, which
is an upper limit on the step energy \cite{oldgrowth}. Since the step energy
is an equilibrium quantity, its calculation using perturbation techniques or
molecular dynamics simulations may be a tractable problem, so this is an
area for further research.

\subsection{The Prefactor $A$}

\begin{figure}[htb] 
  \centering
  \includegraphics[bb=0 0 1631 1263,width=4.8in,height=3.72in,keepaspectratio]{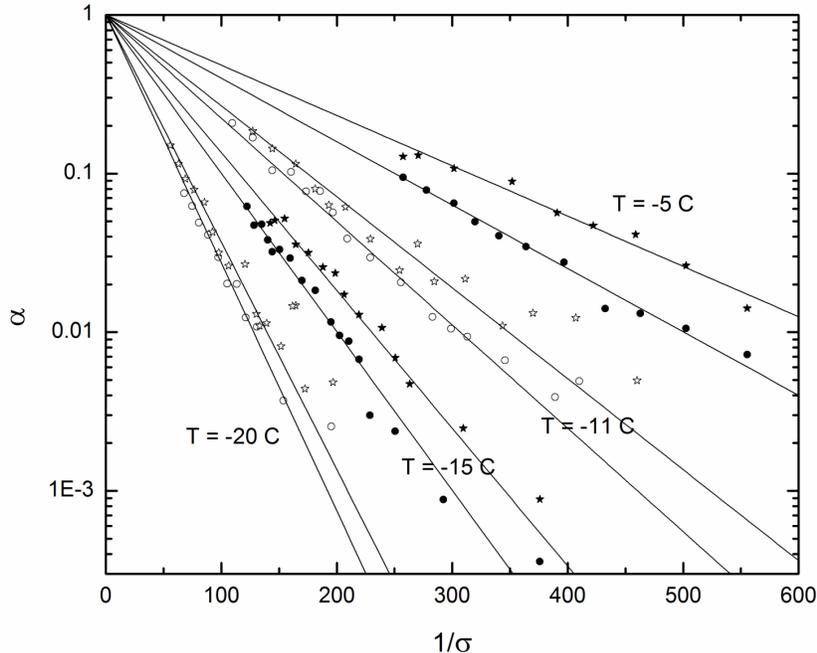}
  \caption{A plot of the measured
attachment coefficient $\protect\alpha $ as a function of the inverse
supersaturation $1/\protect\sigma $ for a representative sample of two
crystals at each of four different temperatures, as labeled in the plot. Fit
lines are drawn to show that the extrapolated data are all consistent with a
prefactor of $A=1,$ as described in the text.}
\label{acomp}
\end{figure}

A remaining question concerns our choice to use a constant prefactor $A$ in
our fitting procedure, and our choice of the numerical value of $A=1.$ We
explored this question at some length by examining the behavior of our data,
and an example is shown in Figure \ref{acomp}. The eight crystals displayed
in this plot were chosen because they exhibited well-behaved growth profiles 
$\alpha (\sigma )$ with minimal growth perturbations from diffusion, and the
residual diffusion effects were fit and corrected before plotting the data.
If the growth satisfies a nucleation model with $\alpha _{intrinsic}=A\exp
(-\sigma _{0}/\sigma ),$ then by plotting the data as shown in Figure \ref%
{acomp}, the extrapolated data should intercept the $\sigma ^{-1}=0$ axis at
the point where $\alpha =A.$ The lines drawn through the points used the
individual $\sigma _{0}$ fit values as described above, which assumed $A=1$.

By examining our data in this fashion, both with and without diffusion
corrections, it became clear that a prefactor of $A=1$ was a good
approximation for the full range of temperatures we measured, at least to an
accuracy of about a factor of two. It was difficult to make an accurate
measurement of $A,$ however, because this involved extrapolating our data,
and such a measurement was somewhat affected by diffusion effects. Including 
$A$ as a free parameter in our fitting procedure for individual crystals
gave the fits too many degrees of freedom and produced unstable results. We
therefore chose to fix $A,$ and we chose $A=1$ from what was essentially a
global fit to the data at all temperatures. As mentioned above, this value
also gives the physically reasonable result that $\alpha
_{intrinsic}\rightarrow 1$ as $\sigma $ becomes large.

To see whether our choice of $A$ had a strong effect on our final results,
we performed an exercise of changing from $A=1$ to $A=0.5$ and reanalyzing
our data. For this we refit each crystal to Equation \ref{fitform} using $%
A=0.5$, generating a new $\sigma _{0}$ and $\alpha _{fit}$ for each crystal,
and from the new data set we recombined the data at each temperature,
following the procedure outlined above. The results are shown in Figure \ref%
{final5}. The points, error bars, and the solid line in this figure are
identical to Figure \ref{finalgraph}, showing the original analysis with $%
A=1.$ The new analysis with $A=0.5$ produced a new set of points shifted
downward, and these are shown as a dashed line in the figure; the error bars
are essentially unchanged from the original analysis.

\begin{figure}[ht] 
  \centering
  \includegraphics[bb=0 0 2798 2233,width=4.3in,height=3.43in,keepaspectratio]{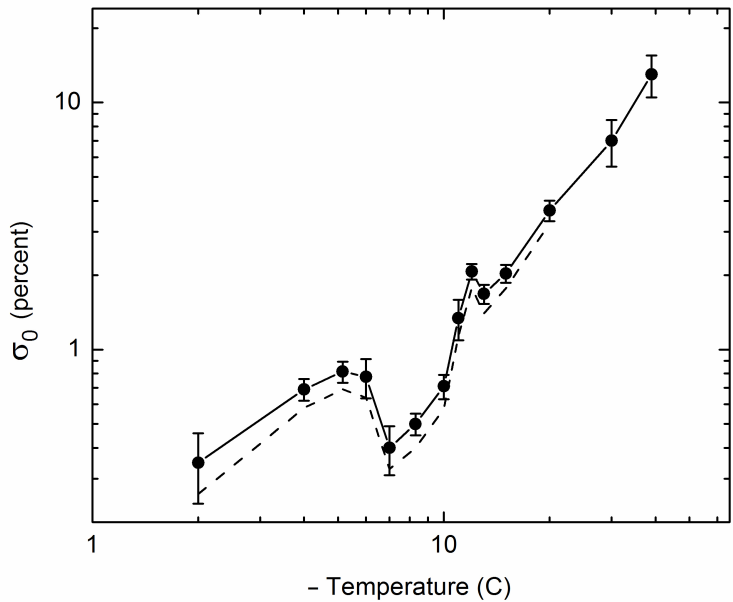}
  \caption{Combined measurements of the
critical supersaturation $\protect\sigma _{0}$ as a function of temperature,
along with a reanalysis using the prefactor $A=0.5.$ The points, error bars,
and solid line are the same as those shown in Figure \protect\ref{finalgraph}%
. The dashed line shows how the results shifted downward when the analysis
was redone using $A=0.5.$}
  \label{final5}
\end{figure}

Although $A=1$ gives a somewhat better global fit to our data, a value of $%
A=0.5$ is not outside the range of potential systematic errors in our
experiment. From our analysis including plots like that shown in Figure \ref%
{acomp}, we believe that $A=1$ is probably accurate to within a factor of
two or better. However we see from Figure \ref{final5} that changing $A$ by
a factor of two produced only a modest shift in our final measurement of $%
\sigma _{0}(T).$

\section{Interpretation and Discussion}

In summary, we measured growth velocities of the (0001) surface of ice as a
function of water vapor supersaturation over the temperature range $-2\leq
T\leq -40.$ From the measured growth velocities we extracted the intrinsic
attachment coefficient $\alpha _{intrinsic}$ using $v=\alpha
_{intrinsic}v_{kin}\sigma _{surf},$ where here $\sigma _{surf}$ is the
supersaturation immediately above the ice surface. Over the entire
temperature range, and for all supersaturations measured, our data were well
described by a simple 2D nucleation model using an attachment coefficient $%
\alpha _{intrinsic}=A\exp (-\sigma _{0}/\sigma ).$ A global fit to the data
yielded the constant prefactor $A=1$, accurate to better than a factor of
two over our entire temperature range. Fits to the data at each temperature
yielded the measured critical supersaturation $\sigma _{0}(T)$ shown in
Figure \ref{finalgraph}.

Referring to Figure \ref{finalgraph}, let us now examine several different
temperature regions and discuss the behavior of $\sigma _{0}(T)$ in detail.

\textbf{Region III: }[$-13>T\geq -40$ C]. In this temperature region we see
a simple power-law behavior in $\sigma _{0}(T),$ which was observed
previously in \cite{oldgrowth}. From the data in Region III we have $\sigma
_{0}(T)\approx 0.009T^{2}$ percent and $\beta (T)\approx 1.2\times
10^{-13}\left\vert T\right\vert $ J/m, where for both these $T$ is in
degrees $C.$

We believe premelting is essentially absent in this temperature region,
meaning that whatever premelting effects do exist are too small to affect
the ice growth dynamics. We have no theoretical explanation for the observed
temperature dependence of $\sigma _{0}(T)$. As the temperature increases,
however, we might expect that surface restructuring would be more likely to
smooth out the edge of a 2D island, thus lowering the step energy.
Quantitative calculation of this effect is an area for further research.

\textbf{The peak at} $T=-12$ C. We examined the small peak in $\sigma _{0}$
at this temperature rather carefully, and it does not appear to be a
statistical fluctuation, nor could we find any systematic effect in our
measurements that would produce a peak at this temperature. The same peak
was reproduced at roughly the same height in three separate runs. We
concluded that the peak is a real feature that probably coincides with the
onset of premelting on the (0001) surface of ice at $T\approx -12$ C. This
onset temperature agrees with that measured by \cite{dosch}, which is
another premelting measurement using a surface preparation similar to that
used in the current experiments. This temperature is also just below the
large dip in $\sigma _{0}$ that we believe indicates a region of partial
premelting (see Region II\ below). It is not obvious why the onset of
premelting would result in a peak in $\sigma _{0},$ however. For this we can
only speculate some unknown many-body phenomenon at the surface.

Our measurements suggest somewhat counterintuitively that ice growth at the
nearby temperatures of $-12$ C and $-13$ C could be qualitatively different
under some circumstances, which could be investigated in other experiments.
Another interesting note is that this is the first time the onset of
premelting has appeared in any measurement as a sharp feature. This suggests
that additional exploration of the detailed growth dynamics of the (0001)
surface of ice near $-12$ C could be fruitful.

\textbf{Region II:} [$-6>T>-11$ C]. In this region we believe that
premelting is partially developed on the (0001) surface. Our intended
meaning here of \textquotedblleft partially developed\textquotedblright\ is
that premelting produces a partially nonfaceted surface that enhances
nucleation and thereby results in the lower values of $\sigma _{0}$
measured, as was previously suggested by Kuroda and Lacmann \cite{kurodalac}%
. Here, as in \cite{kurodalac}, we cannot define a detailed molecular model,
since the surface molecular dynamics must be quite complex at these
temperatures. One's normal picture of a static solid surface with a small
number of admolecules is certainly too simplistic here. Nevertheless, our
data show that the growth is still well fit by a nucleation model with $%
\alpha _{intrinsic}=\exp (-\sigma _{0}/\sigma ).$ We therefore put forth a
simple picture in which a partially developed premelted layer produces a dip
in $\sigma _{0}$ in this temperature region. This picture is supported by
our measurements even if we cannot yet specify a molecular model of the ice
surface dynamics.

\textbf{Region I: [}$-2>T>-5$ C]. At these temperatures we suggest that the
premelted layer has become fully developed, leading to an increase in $%
\sigma _{0}$ compared to that in Region II, again following \cite{kurodalac}%
. Here our picture is that of a solid ice surface covered by a thick
quasi-liquid layer (QLL), with 2D nucleation occurring at the ice/QLL
interface. Because the ice/QLL interface is a smooth, faceted surface, $%
\sigma _{0}$ is higher than that found on a partially premelted surface. To
our knowledge, this is the first time that 2D nucleation at a
solid/quasiliquid interface has been definitively observed.

Concluding, we see that ice crystal growth data such as these can be used
both as a probe of the temperature dependence of premelting and as a measure
of the effects of premelting on ice crystal growth dynamics. We find it
quite remarkable that the growth dynamics can be summed up so concisely by a
single function $\sigma _{0}(T).$ This is true even going through the
transition from no premelting to partially developed premelting to a fully
developed quasiliquid layer, as described above. Although the equilibrium
structure of the ice surface changes dramatically over this temperature
range, as does the equilibrium vapor pressure, the functional form $\alpha
=\exp (-\sigma _{0}/\sigma )$ remains unchanged as the ice growth is
everywhere described by simple 2D nucleation; essentially all that changes
is the critical supersaturation parameter $\sigma _{0}(T).$

This work was supported in part by the California Institute of Technology
and the Caltech-Cambridge Exchange (CamSURF) program.


\begin{thebibliography}{99}
\bibitem{kurodalac} Kuroda, T., and Lacmann, R., \textquotedblleft Growth
Kinetics of Ice from the Vapour Phase and its Growth
Froms,\textquotedblright\ J. Cryst. Growth 56, 189-205 (1982).

\bibitem{kkreview} Kobayashi, T., and Kuroda, T., \textquotedblleft Snow
crystals\textquotedblright , in Morphology of Crystals--Part B, edition I,
Sunagawa (Tokyo: Terra Scientific) 645-743 (1987).

\bibitem{libbrechtreview} Libbrecht, K. G., \textquotedblleft The physics of
snow crystals,\textquotedblright\ Rep. Prog. Phys., 68, 855-895 (2005).

\bibitem{dash} Dash, J. G., Rempel, A. W., and Wettlaufer, J. S.,
\textquotedblleft The physics of premelted ice and its geophysical
consequences,\textquotedblright\ Rev. Mod. Phys. 78, 695-741 (2006)

\bibitem{wei} Wei, X. et al, \textquotedblleft Sum-frequency spectroscopic
studies of ice interfaces,\textquotedblright\ Phys. Rev. B 66 085401 (2002).

\bibitem{dosch} Dosch, H., Lied, A., and Bilgram, J. H., \textquotedblleft
Glancing-angle X-ray scattering studies of the premelting of ice
surfaces,\textquotedblright\ Surf. Sci. 327, 145-164 (1995).

\bibitem{details} Libbrecht, K. G., \textquotedblleft An Improved Apparatus
For Measuring the Growth of Ice Crystals from Water Vapor between -40C and
0C,\textquotedblright\ arXiv:1109.1511 (2011).

\bibitem{taylor} Taylor, J. R., \textquotedblleft An Introduction to Error
Analysis,\textquotedblright\ 2nd Edition, University Science Books (1997).

\bibitem{saito} Saito, Y., \textquotedblleft Statistical Physics of Crystal
Growth,\textquotedblright\ World Scientific Books (1996).

\bibitem{oldgrowth} Libbrecht, K. G., \textquotedblleft Growth Rates of the
Principal Facet of Ice Between -10C and -40C,\textquotedblright\ J. Cryst.
Growth 247, 530 (2003).

\bibitem{precisiongrowth} Libbrecht, K. G.,\textquotedblleft Precision
Measurements of Ice Crystal Growth Rates,\textquotedblright\
arXiv:cond-mat/0608694 (2006).


\end{thebibliography}
\end{document}